\documentstyle[amssymb,12pt]{article}

\begin{document}

\title{DYNAMICAL SYSTEMS AND  POISSON STRUCTURES }
\author{ Metin G{\"{u}}rses$^{1}$, Gusein Sh. Guseinov$^{2}$ and Kostyantyn
Zheltukhin$^{3}$ \\
$^{{\small 1}}${\small Department of Mathematics, Faculty of
Sciences,
Bilkent University,}\\
{\small \ 06800 Ankara, Turkey}\\
{\small gurses@fen.bilkent.edu.tr}\\
$^{{\small 2}}${\small Department of Mathematics, Atilim University,}\\
{\small 06836 Incek, Ankara, Turkey}\\
{\small guseinov@atilim.edu.tr}\\
 $^{{\small 3}}${\small Department of
Mathematics, Middle East Technical
University,}\\
{\small 06531 Ankara, Turkey}\\
{\small  zheltukh@metu.edu.tr}}

\begin{titlepage}
\maketitle

\begin{abstract}
We first consider the Hamiltonian formulation of $n=3$ systems in
general and show that all dynamical systems in  ${\mathbb R}^3$
are bi-Hamiltonian.  An algorithm is introduced to obtain Poisson
structures of a given  dynamical system. We find the Poisson
structures of a dynamical system recently given by Bender et al.
Secondly, we show that all dynamical systems in ${\mathbb R}^n$
are $(n-1)$-Hamiltonian. We give also an algorithm, similar to the
case in ${\mathbb R}^3$, to construct a rank two Poisson structure
of dynamical systems in ${\mathbb R}^n$. We give a classification
of the dynamical systems with respect to the invariant functions
of the vector field $\vec{X}$ and show that all autonomous
dynamical systems in ${\mathbb R}^n$ are super-integrable.
\end{abstract}

\end{titlepage}

\section*{1. Introduction}

Hamiltonian formulation of $n=3$ systems has been intensively
considered in the last two decades. Works \cite{ber1}, \cite{ber2}
on this subject give a very large class of  solutions of the
Jacobi equation for the Poisson matrix $J$. Recently  generalizing
the solutions given in \cite{ber1} we  gave the most general
solution of the Jacobi equation in ${\mathbb R}^3$,  \cite{ay}.
Matrix $J=(J^{ij}), \quad i,j=1,2, \cdots,n$ defines a Poisson
structure in ${\mathbb R}^n$ if it is skew-symmetric,
$J^{ij}=-J^{ji}$, and its entries satisfy the Jacobi equation
\begin{equation}\label{jacobi1}
J^{li}\partial_l\,J^{jk}+J^{lj}\partial_l\,J^{ki}+J^{lk}\partial_l\,J^{ij}=0,
\end{equation}
where $i,j,k=1,2, \cdots,n$. Here we use the summation convention,
meaning that repeated indices are summed up. We showed in
\cite{ay} that the general solution of the above equation
(\ref{jacobi1}) in the case $n=3$ has the  form
\begin{equation}\label{GenSol}
J^{ij}=\mu \epsilon^{ijk}\partial_k  \Psi,~~ i,j=1,2,3,
\end{equation}
where $\mu$ and $\Psi$ are arbitrary differentiable functions of
$x^{i}, t$, $i=1,2,3$ and $\epsilon^{ijk}$ is the Levi-Civita
symbol. Here $t$ should be considered as a parameter. In the same
work we have also considered a bi-Hamiltonian representation of
Hamiltonian systems. It turned out that any Hamiltonian  system in
${\mathbb R}^3$ has a bi-Hamiltonian representation.

In the present paper we prove that any $n$-dimensional dynamical
system
\begin{equation}
\dot{\vec{x}}=\vec{X}(x^{1},x^{2},\ldots ,x^{n},t),
\label{DinSystem}
\end{equation}%
where $\vec{x}=(x^{1},x^{2},\ldots ,x^{n}),$ is Hamiltonian, that
is, has
the form%
\begin{equation}
\dot{{x^{i}}}=J^{ij}\partial _{j}H,~~ i=1,2,\ldots ,n,
\label{HamEqn}
\end{equation}%
where $J=(J^{ij})$ is a Poisson matrix and $H$, as well as
$J^{ij},$ are differentiable functions of the variables
$x_{1},x_{2},\ldots ,x_{n},t$. Moreover, we show that the system
(\ref{DinSystem}) is $(n-1)$-Hamiltonian. This problem in the case
$n=3$ was considered in \cite{haas}, \cite{zhong} where authors
start with an invariant of the dynamical system as a Hamiltonian
and then proceed by writing the system in the form (\ref{HamEqn})
and imposing conditions on $J$ so that it satisfies the Jacobi
equation. But proofs given in these works are, as it seems to us,
incomplete and not satisfactory.

 Using  (\ref{GenSol}) for matrix $J$ we can write equation (\ref{HamEqn}) in ${\mathbb R}^3$  as
\begin{equation}\label{CrossProdEqn}
\dot{\vec{x}}=\mu \vec{\nabla} \Psi \times \vec{\nabla} H.
\end{equation}
Let $\vec X$ be a vector field in ${\mathbb R}^3$. If $H_1$ and
$H_2$ are two invariant functions of  $\vec X$, i.e.,
$\vec{X}(H_\alpha)=X^j\partial_jH_{\alpha}=0, \quad \alpha=1,2$,
then $\vec X$ is parallel to $\vec{\nabla} H_1\times \vec{\nabla}
H_2$. Therefore
\begin{equation}\label{CrossProdField}
\vec{X}=\mu \vec{\nabla} H_1\times \vec{\nabla} H_2,
\end{equation}
where the function $\mu$ is a coefficient of proportionality. The
right-hand side of equation (\ref{CrossProdField}) is in the same
form as the right-hand side of equation (\ref{CrossProdEqn}), so
$\vec{X}$ is a Hamiltonian vector field. We note that the equation
which allows to find the invariants of a vector field $\vec X$ is
a first order linear partial differential equation. We remark here
that dynamical systems in ${\mathbb R}^3$ differ from the
dynamical systems in ${\mathbb R}^n$ for $n>3$. We know the
general solution (\ref{GenSol}) of the Jacobi equation
(\ref{jacobi1})  in ${\mathbb R}^3$. In ${\mathbb R}^n$, as we
shall see in the last section, we know only the rank $2$ solutions
of the Jacobi equations for all $n$.

An important difference of our work, contrary to other works in
the subject, is that in the construction of the Poisson structures
we take into account the invariant functions of the vector field
$\vec{X}$ rather than the invariants (constants of motion) of the
dynamical system. The total time derivative of a differentiable
function $F$ in ${\mathbb R}^n$ along the phase trajectory is
given by

\begin{equation}
{dF \over dt}={\partial F \over \partial t}+\vec{X} \cdot
\vec{\nabla} F.
\end{equation}
An invariant function of the vector field $\vec{X}(x^{1},x^{2},
\ldots,x^{n}, t)$, i.e., $\vec{X} \cdot \vec{\nabla} F=0$, is not
necessarily an invariant function (constant of motion) of the
dynamical system. For autonomous systems where
$\vec{X}=\vec{X}(x^{1},x^{2}, \ldots,x^{n})$ these invariant
functions are the same. We give a representation of the vector
field $\vec{X}$ in terms of its invariant functions. We show that
all autonomous dynamical systems are super-integrable. A key role
plays the existence of $n-1$ functionally independent solutions
$\zeta_{\alpha}(x^{1},x^{2}, \ldots,x^{n},t), (\alpha=1,2,\cdots,
n-1)$ of the linear partial differential equation

\begin{equation}\label{lin1}
\vec{X} \cdot \vec{\nabla} \zeta \equiv X^{1}\, {\partial \zeta
\over
\partial x^{1}}+X^{2}\, {\partial \zeta \over \partial
x^{2}}+\cdots+X^{n}\, {\partial \zeta \over
\partial x^{n}}=0,
\end{equation}
where $X^{i}=X^{i}(x^{1},x^{2},\ldots,x^{n},t)$,
~$i=1,2,\cdots,n$, are given functions (see
\cite{olv}-\cite{sned}). For all $\alpha=1,2,\cdots, n-1$,
$\vec{\nabla} \zeta_{\alpha}$ is perpendicular to the vector field
$\vec{X}$. This leads to the construction of the rank 2 Poisson
tensors for $n>3$:

\begin{equation}\label{poysonn0}
J_{\alpha}^{ij}=\mu \,\epsilon^{\alpha \alpha_{1} \alpha_{2}
\cdots \alpha_{n-2}}\,\,\epsilon ^{ijj_{1}\cdots
j_{n-2}}\,\partial _{j_{1}}\zeta_{\alpha_{1}}\,\partial
_{j_{2}}\,\zeta_{\alpha_{2}}\cdots
\partial _{j_{n-2}}\zeta_{\alpha_{n-2}},
\end{equation}
where $i,j=1,2,\cdots ,n$, and $\alpha=1,2,\cdots, n-1$. Here
$\epsilon ^{ijj_{1}\cdots j_{n-2}}$ and $\epsilon^{\alpha
\alpha_{1} \alpha_{2} \cdots \alpha_{n-2}}$ are Levi-Civita
symbols in $n$ and $n-1$ dimensions respectively. Any dynamical
system with the vector field $\vec{X}$ possesses  Poisson
structures in the form given in (\ref{poysonn0}). Hence we can
give a classification of dynamical systems in ${\mathbb R}^n$ with
respect to the invariant functions of the vector field $\vec{X}$.
There are mainly three classes where the super-integrable
dynamical systems constitute the first class. By the use of the
invariant functions of the vector field $\vec{X}(x^{1},x^{2},
\ldots,x^{n}, t)$ in general we give a Poisson structure in
${\mathbb R}^n$ which has rank 2. For autonomous systems, the form
(\ref{poysonn0}) of the above Poisson structure first was given in
the works \cite{raz} and \cite{nam}.

Our results in this work are mainly local. This means that our
results are valid in an open domain of  ${\mathbb R}^n$ where the
Poisson structures are different from zero. In \cite{ay} we showed
that the Poisson structure (\ref{GenSol}) in ${\mathbb R}^3$
preserves its form in the neighborhood of irregular points, lines
and planes.

In the next section we give new  proofs of the formula
(\ref{GenSol}) and prove that any dynamical system in ${\mathbb
R}^3$ is Hamiltonian. So, following \cite{ay} we show that any
dynamical system in ${\mathbb R}^3$ is bi-Hamiltonian.
Applications of these theorems to several dynamical systems are
presented. Here we also show that the dynamical system given by
Bender at al \cite{ben} is bi-Hamiltonian. In section 3 we discuss
Poisson structures in ${\mathbb R}^n$. We give a representation of
the Poisson structure in ${\mathbb R}^n$ in terms of the invariant
functions of the vector field $\vec{X}$. Such a representation
leads to a classification of dynamical systems with respect to
these functions.

\section*{2. Dynamical Systems in ${\mathbb R}^3$}

Although the proof of (\ref{GenSol}) was given in \cite{ay}, here
we shall give two simpler proofs. The first one is a shorter proof
than the one given in \cite{ay}. In the sequel we use the
notations $x^{1}=x, x^2=y, x^3=z$.

\vspace{0.3cm}

 \noindent
 {\bf Theorem 1.} {\it All
Poisson structures in ${\mathbb R}^3$ have the form
(\ref{GenSol}), i.e., $J^{ij}=\mu\, \epsilon^{ijk}\,
\partial_{k}\, H_{0}$. Here $\mu$ and $H_{0}$ are some
differentiable functions of $x^{i}$ and $t$, ($i=1,2,3$) }

\vspace{0.3cm}

\noindent
{\bf Proof.}
Any skew-symmetric second rank tensors in
${\mathbb R}^3$ can be given as
\begin{equation}\label{j1}
J^{ij}=\epsilon^{ijk} J_{k}, ~~ i,j=1,2,3,
\end{equation}
where $J_{1},J_{2}$ and $J_{3}$ are differentiable functions in
${\mathbb R}^3$ and we assume that there exists a domain $\Omega$
in ${\mathbb R}^3$ so that these functions do not vanish
simultaneously. When (\ref{j1}) inserted into the Jacobi equation
(\ref{jacobi1}) we get
\begin{equation}\label{j2}
\vec{J} \cdot (\vec{\nabla} \times \vec{J})=0,
\end{equation}
where $\vec{J}=(J_{1},J_{2},J_{3})$ is a differentiable vector
field in ${\mathbb R}^3$ not vanishing in $\Omega$. We call
$\vec{J}$ as the Poisson vector field. It is easy to show that
(\ref{j2}) has a local scale invariance. Let $\vec{J}=\psi\,
\vec{E}$, where $\psi$ is an arbitrary function. If  $\vec{E}$
satisfies (\ref{j2}) then $\vec{J}$ satisfies the same equation.
Hence it is enough to show that $\vec{E}$ is proportional to the
gradient of a function. Using freedom of local scale invariance we
can take $\vec{E}=(u,v,1)$ where $u$ and $v$ are arbitrary
functions in ${\mathbb R}^3$. Then (\ref{j2}) for vector $\vec{E}$
reduces to
\begin{equation}\label{j3}
\partial_y u-\partial_x v-v \partial_z u+u \partial_z v=0,
\end{equation}
where $x,y,z $ are local coordinates.
Letting $u={\partial_x f \over \rho}$ and $v={\partial_y f \over \rho}$, where
$f$ and $\rho$ are functions of $x,y,z$ we get
\begin{equation}
\partial_x f\, \partial_y(\rho-\partial_z f)-\partial_y f\partial_x(\rho-\partial_z
f)=0.
\end{equation}
General solution of this equation is given by
\begin{equation}
\rho-\partial_z f=h(f,z),
\end{equation}
where $h$ is an arbitrary function of $f$ and $z$. Then the vector
filed $\vec{E}$ takes the form
\begin{equation}\label{j4}
\vec{E}={1 \over \partial_z f+h}\, (\partial_x f,\partial_y
f,\partial_z f+h).
\end{equation}
Let  $g(f,z)$ be a function satisfying $g_{,z}=h\partial_f g$.
Here we note that $\partial_{z}\, g(f,z)={\partial g \over
\partial f}\, {\partial_{z} f}+g_{,z}$ where
$g_{,z}=\partial_{s}\,g(f(x,y,z),s)|_{s=z}$. Then (\ref{j4})
becomes
\begin{equation}
\vec{E}={1 \over (\partial_zf+h) \partial_f g}\, \vec{\nabla}\, g,
\end{equation}
which completes the proof. Here $\partial_{f}\, g={\partial g
\over
\partial f}$. $\Box$

\vspace{0.3cm} \noindent The second proof is an indirect one which
is given in \cite{sned} (Theorem 5 in this reference).

\vspace{0.3cm} \noindent {\bf Definition 2}.\, {\it Let $\vec F$
be a vector field in ${\mathbb R}^3$. Then the equation $ \vec{F}
\cdot d \vec{x}=0$ is called a Pfaffian differential equation. A
Pfaffian  differential equation is called integrable if the 1-form
$\vec{F} \cdot d \vec{x}=\mu d H $, where $\mu$ and $H$ are some
differentiable functions in ${\mathbb R}^3$.}

\vspace{0.3cm}

\noindent
 Let us now consider the Pfaffian differential equation
with the Poisson vector field $\vec{J}$ in (\ref{j1})
\begin{equation}
\vec{J} \cdot d\vec{x}=0.
\end{equation}
For such  Pfaffian differential equations we have the following
result (see \cite{sned}).

 \vspace{0.3cm}

\noindent {\bf Theorem 3}. {\it A necessary and sufficient
condition that the Pfaffian differential equation $\vec{J} \cdot d
\vec{x}=0$ should be integrable is that $\vec{J} \cdot
(\vec{\nabla} \times \vec{J})=0$.}

\vspace{0.3cm} \noindent By (\ref{j2}), this theorem implies that
$\vec{J}=\mu {\vec \nabla} \Psi$.

A well known example of a dynamical system with Hamiltonian
structure of the form (\ref{HamEqn}) is the Euler equations.

\vspace{0.3cm}

\noindent {\bf Example 1.} The Euler equations \cite{olv} are
\begin{equation}\label{e}
\begin{array}{lll}
\dot{x}&=&\displaystyle{\frac{I_2-I_3}{I_2I_3}}yz,\\
\dot{y}&=&\displaystyle{\frac{I_3-I_1}{I_3I_1}}xz,\\
\dot{z}&=&\displaystyle{\frac{I_1-I_2}{I_1I_2}}xy,\\
\end{array}
\end{equation}
where $ I_1,I_2,I_3\in {\mathbb R}$ are some (non-vanishing) real
constants. This system admits Hamiltonian representation of the
form~(\ref{HamEqn}). The matrix $J$ can be defined in terms of
functions $\Psi=H_{0}=-\frac{1}{2}(x^2+y^2+z^2)$ and $\mu=1$, and
we take $H=H_{1}=\displaystyle{\frac{x^2}{2I_1}+\frac{y^2}{2I_2}+\frac{z^2}{2I_3}}$.\\

Writing the Poisson structure in the form
(\ref{GenSol}) allows us to construct bi-Hamiltonian
representations of a given Hamiltonian system.

\vspace{0.3cm}

\noindent {\bf Definition 4.} {Two Poisson structures $J_{0}$ and
$J_{1}$ are compatible, if the sum $J_{0}+J_{1}$ defines also a
Poisson structure.}

\vspace{0.3cm}

\noindent {\bf Lemma 5.} {\it Let $\mu, H_{0},$ and $H_{1}$ be
arbitrary differentiable functions. Then the Poisson  structures
$J_{0}$ and $J_{1}$ given by
$J_{0}^{ij}=\mu\epsilon^{ijk}\partial_k \,H_{0}$  and
$J_{1}^{ij}=-\mu \epsilon^{ijk}\partial_k \,H_{1}$ are
compatible.}

\vspace{0.3cm}

\noindent This suggests that all Poisson structures in ${\mathbb
R}^3$ have compatible companions. Such compatible Poisson
structures can be used to construct bi-Hamiltonian systems (for
Hamiltonian and bi-Hamiltonian systems see \cite{olv},\cite{Blz}
and the references therein).

\vspace{0.3cm}

\noindent {\bf Definition 6.} {\it A Hamiltonian equation is said
to be bi-Hamiltonian if it admits compatible Poisson structures
$J_{0}$ and $J_{1}$ with the corresponding Hamiltonian functions
$H_{1}$ and $H_{0}$ respectively, such that
\begin{equation}
\frac{dx}{dt}=J_{0}\nabla H_{1}=J_{1}\nabla\,H_{0}.
\end{equation}
}
 \vspace{0.3cm}

\noindent {\bf Lemma 7.} {\it Let $J_{0}$ be given by
(\ref{GenSol}), i.e., $ J_{0}^{ij}=\mu \epsilon ^{ijk}\partial
_{k}\,H_{0},$  and let $H_{1}$ be any differentiable function,
then the Hamiltonian equation
\begin{equation}
\frac{dx}{dt}=J_{0}\nabla H_{1}=J_{1}\nabla\,H_{0} =\mu\,
{\vec{\nabla} H_{1}} \times {\vec{\nabla} H_{0}},
\end{equation}
is bi-Hamiltonian with the second Poisson structure given by
$J_{1}$ with entries $J_{1}^{ij}=-\mu \epsilon^{ijk}\partial_k
H_{1}$ and the second Hamiltonian $H_{0}$.}

\vspace{0.3cm}

Let us prove that any dynamical system in ${\mathbb R}^3$ has
Hamiltonian form.

\vspace{0.3cm}

\noindent
 {\bf Theorem 8}.\,{\it All dynamical systems in ${\mathbb R}^3$
 are Hamiltonian. This means that any vector field $\vec{X}$ in ${\mathbb R}^3$ is
Hamiltonian vector  field. Furthermore all
 dynamical systems in ${\mathbb R}^3$ are bi-Hamiltonian.}

 \vspace{0.3cm}

 \noindent
 {\bf Proof}. Let $\zeta$ be an invariant function of the vector
 field $\vec{X}$, i.e., $X(\zeta) \equiv \vec{X} \cdot \vec{\nabla}
 \zeta=0$. This gives a first order linear differential equation in ${\mathbb R}^3$
 for $\zeta$. For a given vector field $\vec{X}=(f,g,h)$ this
 equation becomes
\begin{equation}\label{first}
f(x,y,z,t)\, {\partial \zeta \over \partial x}+ g(x,y,z,t)\,
{\partial \zeta \over
\partial y}+ h(x,y,z,t)\, {\partial \zeta \over \partial z}=0,
\end{equation}
where $x, y, z$ are local coordinates. From the theory of first
order linear partial differential equations \cite{olv},
\cite{olv1}, \cite{sned} the general solution of this partial
differential equation can be determined from the following set of
equations

\begin{equation}\label{syst}
{dx \over f(x,y,z,t)}={dy \over g(x,y,z,t)}={dz \over h(x,y,z,t)}.
\end{equation}
There exist two functionally independent solutions $\zeta_{1}$ and
$\zeta_{2}$ of (\ref{syst}) in an open domain $D \subset {\mathbb
R}^3$ and the general solution of (\ref{first}) will be an
arbitrary function of $\zeta_{1} $ and $\zeta_{2}$, i.e.,
$\zeta=F(\zeta_{1}, \zeta_{2})$. This implies that the vector
field $\vec{X}$ will be orthogonal to both $\vec{\nabla}
\zeta_{1}$ and $\vec{\nabla} \zeta_{2}$. Then $\vec{X}=\mu \,
(\vec{\nabla} \zeta_{1}) \times (\vec{\nabla} \zeta_{2})$. Hence
the vector field $\vec{X}$ is Hamiltonian by (\ref{CrossProdEqn}).
$\Box$

\vspace{0.3cm}

 \noindent
 This theorem gives also an algorithm to find the Poisson
 structures or the functions $H_{0}$, $H_{1}$ and $\mu$ of a given
 dynamical system. The functions $H_{0}$ and $H_{1}$ are the
 invariant functions of the vector  field $\vec{X}$ which
 can be determined by solving the system equations (\ref{syst})
 and $\mu$ is determined from
 \begin{equation}\label{mu}
 \mu={\vec{X} \cdot \vec{X} \over \vec{X} \cdot (\vec{\nabla}\,
 H_{0} \times \vec{\nabla}\, H_{1})}.
 \end{equation}
 Note that $\mu$ can also be determined from
 \begin{eqnarray}\label{mu1}
 \mu&=&{X^{1} \over \partial_{2} H_{0} \partial_{3} H_{1}-\partial_{3} H_{0} \partial_{2} H_{1}}\\
 &=& {X^{2} \over \partial_{3} H_{1} \partial_{3} H_{1}-\partial_{1} H_{0} \partial_{3} H_{1}} \nonumber\\
 &=& {X^{3} \over \partial_{1} H_{0} \partial_{2} H_{1}-\partial_{2} H_{0}
 \partial_{1}H_{1}}. \nonumber
 \end{eqnarray}

\vspace{0.3cm}

\noindent {\bf Example 2.} As an application of the method
described above we consider Kermac-Mckendric system
\begin{equation}
\begin{array}{lll}
\dot{x}&=&-rxy,\\
\dot{y}&=&rxy-ay,\\
\dot{z}&=&ay,\\
\end{array}
\end{equation}
where $r,a\in {\mathbb R}$ are constants. Let us put the system
into Hamiltonian form. For the  Kermac-Mckendric system, equations
(\ref{syst}) become
\begin{equation}\label{syst1}
{dx \over -r xy}={dy \over rxy-a y}={dz \over ay}.
\end{equation}
Here $a$ and $r$ may depend on $t$ in general. Adding the
numerators and denominators of (\ref{syst1}) we get
\begin{equation}
{dx \over -r xy}={dx+dy+dz \over 0}.
\end{equation}
Hence $H_{1}=x+y+z$ is one of the invariant functions of the
vector field. Using the first and last terms in (\ref{syst1}) we
get
\begin{equation}
{dx \over -r x}={dz \over a},
\end{equation}
which gives $H_{0}=r\,z+a\ln x$ as the second invariant function
of the vector field $\vec{X}$. Using (\ref{mu}) we get $\mu=xy$.
Since $\vec X=\mu \vec{\nabla} H_0 \times \vec{\nabla} H_1$, the
system admits a Hamiltonian representation where the Poisson
structure $J$ is given by (\ref{GenSol}) with $\mu=xy$,
$\Psi=H_{0}=rz+a\ln x$, and the Hamiltonian is $H_{1}=x+y+z$.

\vspace{0.3cm}

\noindent {\bf Example 3.} The dynamical system is given by
\begin{equation}
\begin{array}{lll}
\dot{x}&=&yz (1+2x^2\, N/D),\\
\dot{y}&=&-2xz (1-y^2 N/D),\\
\dot{z}&=&xy(1+2 z^2 N/D),\\
\end{array}
\end{equation}
where $N=x^2+y^2+z^2-1,~D=x^2 y^2+y^2 z^2+4 x^2 z^2$. This example
was obtained by Bender et
 all \cite{ben} by complexifying the Euler system in
 Example 1. They claim that this system is not Hamiltonian
 apparently bearing in mind the more classical definition of a Hamiltonian
system. Using the Definition 6 we  show that this system is not
only Hamiltonian but also bi-Hamiltonian. We obtain that

\begin{equation}
H_{0}={(N+1)^2 \over D}\,N,~~ H_{1}={x^2-z^2 \over D}(2y^2 z^2+4
x^2 z^2 +y^4+2x^2 y^2-y^2).
\end{equation}
Here
\begin{equation}
\mu={D^2 \over 4[3D^2+D\, P+Q]},
\end{equation}
where
\begin{eqnarray}
P&=&-2x^4+4y^4-4x^2 y^2+x^2-2y^2-4y^2 z^2+14 z^4+z^2, \nonumber \\
Q&=&-2x^8+12x^6 z^2+2 x^6-20 x^4 z^4 -6x^4 z^2-52 x^2 z^6-6 x^2
z^4 \nonumber\\
&& +y^8-y^6+4y^4 z^4-16 y^2 z^6 -2 z^8 +2 z^6.
\end{eqnarray}
Indeed these invariant functions were given in \cite{ben} as
functions $A$ and $B$. The reason why Bender et al \cite{ben}
concluded that the system in Example 3 is non-Hamiltonian is that
the vector filed $\vec{X}$ has nonzero divergence. It follows from
$\vec{X}=\mu \vec{\nabla}H_{0}\times \vec{\nabla}%
H_{1}$ that $\vec{\nabla}\cdot \left( {\frac{1}{\mu
}}\,\vec{X}\right) =0$. When $\mu $ is not a constant the
corresponding Hamiltonian vector field has a nonzero divergence.

\vspace{0.3cm}

\noindent {\bf Remark 1}.\, \thinspace\ With respect to the time
dependency of
invariant functions of the vector field $\vec{X}$ dynamical systems in $%
{\Bbb R}^{3}$ can be split into three classes.

\vspace{0.3cm}

\noindent {\bf Class A}.\, Both invariant functions $H_{0}$ and
$H_{1}$ of the vector field $\vec{X}$ do not depend on time
explicitly. In this case both $H_{0}$ and $H_{1}$ are also
invariant functions of the dynamical systems. Hence the system is
super-integrable. All autonomous dynamical systems such as the
Euler equation (Example 1) and the Kermac-Mckendric system
(Example 2) belong to this class.

\vspace{0.3cm}

\noindent {\bf Class B}.\, One of the invariant functions $H_{0}$
and $H_{1}$ of the vector field $\vec{X}$ depends on $t$
explicitly. Hence the other one is an invariant function also of
the dynamical system. When $I_{1}, I_{2}$ and $I_{3}$ in Example 1
are time dependent the Euler system becomes the member of this
class. In this case $H_{0}$ is the Hamiltonian function and
$H_{1}$ is the function defining the Poisson structure. Similarly,
in Example 2 we may consider the parameters $a$ and $r$ as time
dependent. Then Kermac-Mckendric system becomes also a member of
this class.

\vspace{0.3cm}

\noindent {\bf Class C}.\, Both $H_{0}$ and $H_{1}$ are explicit
functions of time variable $t$ but they are not the invariants of
the system. There may be invariants of the dynamical system. Let
$F$ be such an invariant. Then

\begin{equation}
{dF \over dt} \equiv {\partial F \over \partial
t}+\{F,H_{1}\}_{0}={\partial F \over \partial
t}+\{F,H_{0}\}_{1}=0,
\end{equation}
where for any $F$ and $G$
\begin{equation}
\{F,G\}_{\alpha} \equiv J_{\alpha}^{ij}\, \partial_{i} \,F
\partial_{j}\, G,~~~ \alpha=0,1.
\end{equation}

\vspace{0.3cm}

\section*{3. Poisson structures in ${\mathbb R}^n$}

Let us consider the dynamical system

\begin{equation}
{\frac{dx^{i}}{dt}}=X^{i}(x^{1},x^{2},\cdots
,x^{n},t),~~i=1,2,\cdots ,n. \label{dyn3}
\end{equation}

\bigskip

\noindent {\bf Theorem 9. }{\it All dynamical systems in }${\Bbb
R}^{n}$ {\it are Hamiltonian. Furthermore all dynamical systems in
}${\Bbb R}^{n}$ {\it are} $(n-1)${\it -Hamiltonian.}

\noindent {\bf Proof.} Extending the proof of Theorem 8 to ${\Bbb
R} ^{n}$ consider the linear partial differential equation
(\ref{lin1}). There exist $n-1$ functionally independent solutions
$H_{\alpha}, (\alpha=1,2, \cdots, n-1)$ of this equation (which
are invariant functions of the vector field $\vec{X}$)
\cite{olv}-\cite{sned}. Since $\vec{X}$ is orthogonal to the vectors $\vec{\nabla}%
H_{\alpha },~(\alpha =1,2,\cdots ,n-1),$ we have

\begin{equation}
\vec{X}= \mu \left|
\begin{array}{rrrrrr}
\vec{e_{1}}& \vec{e_{2}} & \cdot& \cdot& \cdot &\vec{e_{n}}\\
\partial_{1} H_{1}&\partial_{2} H_{1}& \cdot&\cdot&\cdot&\partial_{n} H_{1} \\
\cdot& \cdot&\cdot&\cdot&\cdot&\cdot\\
\cdot& \cdot&\cdot&\cdot&\cdot&\cdot\\
\partial_{1} H_{n-1}&\partial_{2} H_{n-1}& \cdot&\cdot&\cdot&\partial_{n} H_{n-1}
\end{array}
\right|,
\end{equation}
where the function $\mu$ is a coefficient of proportionality and
$\vec{e_{i}}$ is $n$-dimensional unit vector with the $i$th
coordinate $1$ and remaining coordinates $0$. Therefore
\begin{equation}
X^{i}=\mu \epsilon ^{ij_{1}j_{2}\cdots j_{n-1}}\partial
_{j_{1}}H_{1}\partial _{j_{2}}H_{2}\cdots \partial
_{j_{n-1}}H_{n-1}.
\end{equation}
Hence all dynamical systems (\ref{dyn3}) have the Hamiltonian
representation
\begin{equation}\label{sysn}
{\frac{dx^{i}}{dt}}=J_{\alpha}^{ij}\partial
_{j}H_{\alpha},~~i=1,2,\cdots ,n, \label{dyn4}~~(\mbox{no sum
on}~~ \alpha)
\end{equation}
with
\begin{equation}\label{poysonn}
J_{\alpha}^{ij}=\mu \epsilon^{\alpha \alpha_{1} \alpha_{2} \cdots
\alpha_{n-2}}\,\epsilon ^{ijj_{1}\cdots j_{n-2}}\,\partial
_{j_{1}}H_{\alpha_{1}}\,\partial _{j_{2}}\,H_{\alpha_{2}} \cdots
\partial _{j_{n-2}}H_{\alpha_{n-2}},
\end{equation}
where $i,j=1,2,\cdots ,n$, $\alpha=1,2,\cdots,n-1$. Here $\epsilon
^{ijj_{1}\cdots j_{n-2}}$ and $\epsilon^{\alpha \alpha_{1}
\alpha_{2} \cdots \alpha_{n-2}}$ are Levi-Civita symbols in $n$
and $n-1$ dimensions respectively. The function $\mu$ can be
determined, for example, from

\begin{equation}
\mu={X^{1} \over \left|
\begin{array}{rrrrr}
\partial_{2} H_{1}& \cdot&\cdot&\cdot&\partial_{n} H_{1} \\
\cdot& \cdot&\cdot&\cdot&\cdot\\
\cdot& \cdot&\cdot&\cdot&\cdot\\
\partial_{2} H_{n-1}& \cdot&\cdot&\cdot&\partial_{n} H_{n-1}
\end{array}
\right|.  }
\end{equation}
It can be seen that the matrix $J_{\alpha}$ with the entries
$J_{\alpha}^{ij}$ given by (\ref{poysonn}) defines a Poisson
structure in ${\mathbb R}^n$ and since
\begin{equation}
J_{\alpha} \cdot \nabla H_{\beta}=0, ~~\alpha, \beta=1,2, \cdots,
n-1,
\end{equation}
with $\beta \ne \alpha$, the rank of the matrix $J_{\alpha}$
equals 2 (for all $\alpha=1,2,\cdots, n-1$). In (\ref{sysn}) we
can take  any of $H_{1}, H_{2}, \cdots, H_{n-1}$ as the Hamilton
function and use the remaining
$H_{k}$'s in (\ref{poysonn}). We observe that all dynamical systems (\ref{dyn3}) in ${\Bbb R}^{n}$ have $%
n-1$ number of different Poisson structures in the form given by
(\ref{poysonn}). The same system may have a Poisson structure with
a rank higher than two. The following example clarifies this
point.

\vspace{0.3cm}

\noindent {\bf Example 4}. Let
\begin{equation}
{\dot{x_{1}}}=x_{4},~{\dot{x_{2}}}=x_{3},~{\dot{x_{3}}}=-x_{2},~{\dot{x_{4}}}=-x_{1}.
\end{equation}
Clearly this system admits a Poisson structure with rank four
\begin{equation}
J= \left(
\begin{array}{rrrr}
0& 0 & 0&1\\
0&0&1&0 \\
0& -1&0&0\\
-1&0&0&0
\end{array}
\right), ~~ H={1 \over 2}(x_{1}^2+x_{2}^2+x_{3}^2+x_{4}^2).
\end{equation}
The invariant functions of the vector field
$\vec{X}=(x_{4},x_{3},-x_{2},-x_{1})$ are
\begin{eqnarray}
H_{1}&=&{1 \over 2}(x_{1}^2+x_{2}^2+x_{3}^2+x_{4}^2),\\
H_{2}&=&{1 \over 2}(x_{2}^2+x_{3}^2),\\
H_{3}&=&x_{1}\,x_{3}-x_{2}\,x_{4}.
\end{eqnarray}
Then the above system has three different ways of representation
with the second rank Poisson structures
\begin{eqnarray}
J_{1}^{ij}&=&\mu \epsilon^{ijkl}\, \partial_{k}\, H_{1}
\partial_{l}
H_{2},~~ H=H_{3},\\
J_{2}^{ij}&=&-\mu \epsilon^{ijkl}\, \partial_{k}\, H_{1}
\partial_{l}
H_{3},~~ H=H_{2},\\
J_{3}^{ij}&=&\mu \epsilon^{ijkl}\, \partial_{k}\, H_{2}
\partial_{l}
H_{3},~~ H=H_{1},
\end{eqnarray}
where $\mu (x_{1}x_{2}+x_{3}x_{4})=1$. These Poisson structures
are compatible not only pairwise but also triple-wise. This means
that any linear combination of these structures is also a Poisson
structure. Let $J=\alpha_{1} J_{1}+\alpha_{2} J_{2}+\alpha_{3}
J_{3}$ then it is possible to show that
\begin{equation}
J^{ij}=\mu \epsilon^{ijkl}\, \partial_{k}\, {\tilde H}_{1}
\partial_{l}
{\tilde H}_{2},
\end{equation}
where $\tilde{H}_{1}$ and $\tilde{H}_{2}$ are linear combinations
of $H_{1}, H_{2}$ and $H_{3}$,
\begin{eqnarray}
\tilde{H}_{1}&=&H_{1}-{\alpha_{3} \over \alpha_{2}} H_{2},~
\tilde{H}_{2}=\alpha_{1} H_{2}-\alpha_{2} H_{3} ~~\mbox{if}
~~\alpha_{2} \ne 0,\\
\tilde{H}_{1}&=&\alpha_{1} H_{1}-\alpha_{3}  H_{2},~
\tilde{H}_{2}= H_{2} ~~\mbox{if}~~ \alpha_{2} = 0.
\end{eqnarray}

\vspace{0.3cm} \noindent {\bf Definition 10}. {\it  A dynamical
system (\ref{dyn3}) in ${\mathbb R}^n$ is called super-integrable
if it has $n-1$ functionally independent first integrals
(constants of motion)}.

\vspace{0.3cm}

\noindent {\bf Theorem 11}. {\it All autonomous dynamical systems
in ${\mathbb R}^n$ are super-integrable}.

\vspace{0.3cm}

\noindent
{\bf Proof}. If the system (\ref{dyn3}) is autonomous,
then the vector field $\vec{X}$ does not depend on $t$ explicitly.
Therefore each of the invariant functions
$H_{\alpha},~(\alpha=1,2,\cdots, n-1)$ of the vector field
$\vec{X}$ is a constant of motion of the system (\ref{dyn3}).

\vspace{0.3cm} \vspace{0.3cm} Some (or all) of the invariant
functions $ H_{\alpha }$, $(\alpha =1,2,\cdots ,n-1)$ of the
vector field $\vec{X}$ may depend on $t$. Like in ${\mathbb R}^3$
we can classify the dynamical systems in ${\mathbb R}^n$ with
respect to the invariant functions of the vector field
$\vec{X}(x^{1},x^{2}, \cdots,x^{n}, t)$.

\vspace{0.4cm}

\noindent {\bf Class A}.\, All invariant functions $H_{\alpha},~
(\alpha=1,2,\cdots. n-1)$ of the vector field $\vec{X}$ do not
depend on $t$ explicitly. In this case all functions $H_{\alpha},~
(\alpha=1,2,\cdots. n-1)$ are also invariant functions (constants
of motion) of the dynamical system. Hence the system is
super-integrable. In the context of the the multi- Hamiltonian
structure, such systems were first studied by \cite{nam} and
\cite{raz}. The form (\ref{poysonn}) of the Poisson structure was
given in these works.  Its properties were investigated in
\cite{nut}.

\vspace{0.3cm}

\noindent {\bf Class B}.\, At least one of the invariant functions
$H_{\alpha},~ (\alpha=1,2,\cdots. n-1)$ of the vector field
$\vec{X}$ does not depend on $t$ explicitly. That function is  an
invariant function also of the dynamical system.

\vspace{0.3cm}

\noindent {\bf Class C}.\, All $H_{\alpha},~ (\alpha=1,2,\cdots.
n-1)$ are explicit function of time variable $t$ but they are not
the invariants of the system. There may be invariants of the
dynamical system. Let $F$ be such an invariant. Then

\begin{equation}
{dF \over dt} \equiv {\partial F \over \partial
t}+\{F,H_{\alpha}\}_{\alpha}=0,~~\alpha=1,2,\cdots, n-1
\end{equation},
where for any $F$ and $G$
\begin{equation}
\{F,G\}_{\alpha} \equiv J_{\alpha}^{ij}\, \partial_{i} \,F
\partial_{j}\, G,~~~ \alpha=0,1,\cdots, n-1.
\end{equation}

 \vspace{1cm}

 \noindent {\bf Acknowledgements}: \\
 We wish to thank Prof. M. Blaszak for
critical reading of the paper and for constructive comments. This
work is partially supported by the Turkish Academy of Sciences and
by the Scientific and Technical Research Council of Turkey.

\end{document}